\documentstyle[twocolumn,prl,aps,epsf]{revtex}
\begin{document}

\title{Nuclear structure studies with the $^{7}$Li$(e,e^{\prime} p)$
reaction}

\author{L.~Lapik\'as$^1$, J.~Wesseling$^1$ and R.B.~Wiringa$^2$}

\address{
$^1$ NIKHEF, P.O.~Box~41882, 1009~DB~Amsterdam, The~Netherlands\\ $^2$
Physics Division, Argonne National Laboratory, Argonne, IL 60439, USA
}

\date{March 10, 1999}

\maketitle

\begin{abstract}
Experimental momentum distributions for the transitions to the ground
state and first excited state of $^{6}$He have been measured via the
reaction $^{7}$Li($e,e'p$)$^{6}$He, in the missing momentum range from
-70 to 260 MeV/c.  They are compared to theoretical distributions
calculated with mean-field wave functions and with variational Monte
Carlo (VMC) wave functions which include strong state-dependent
correlations in both $^{7}$Li and $^{6}$He.  These VMC calculations
provide a parameter-free prediction of the momentum distribution that
reproduces the measured data, including its normalization.  The deduced
summed spectroscopic factor for the two transitions is 0.58$\pm$0.05,
in perfect agreement with the VMC value of 0.60.  This is the first
successful comparison of experiment and {\it ab initio} theory for
spectroscopic factors in p-shell nuclei.

\vspace{0.4 cm}
\vspace{0.4 cm}

\end{abstract}



In this letter we present new experimental data on the reaction
$^{7}$Li($e,e'p$)$^{6}$He and compare the deduced momentum
distributions with recent {\it ab initio} predictions from variational
Monte Carlo (VMC) calculations.
Some years ago, two of us (Lapik\'as and Wesseling) participated in
electron scattering experiments to determine shell occupancies in the
nuclei $^{30}$Si, $^{31}$P, and $^{32}$S \cite{Wes92,Wes97}.
In the latter case, a Li$_{2}$S target was used, with resolution
sufficient to separate the discrete transitions in the reaction
$^{32}$S($e,e'p$)$^{31}$P from those in the reaction
$^{7}$Li($e,e'p$)$^{6}$He.
However, the results for $^{7}$Li were not published at the time.
Independently, one of us (Wiringa) recently calculated the overlap wave
function $\langle^{6}$He$\mid~a({\bf r})\mid^{7}$Li$\rangle$ as part of
a general program of quantum Monte Carlo studies of the light p-shell
nuclei \cite{Pud97,Wir98}.
These calculations use realistic two- and three-nucleon interactions
fit to NN scattering data and few-body nuclear bound states, and
produce fully correlated $A$-body wave functions.
We have now found that using the VMC overlap as input to a Coulomb
Distorted Wave Impulse Approximation (CDWIA) code results in excellent
predictions of the observed momentum distributions and transition rms
radii {\it including the absolute normalization} of the cross sections.
This is the first successful comparison of experiment and {\it ab
initio} theory for spectroscopic factors in p-shell nuclei.

The effect of including short-range and tensor correlations in the
calculation of nuclear structure has been studied previously in detail
for few-body systems.  In particular, momentum distributions for the
nuclei $^{2}$H \cite{Blo98,Kas98}, $^{3}$He \cite{Mar88}, and $^{4}$He
\cite{Gof94,Lee98}, measured via the reaction ($e,e'p$), have been
compared to calculations (Faddeev \cite{Gol95} and VMC
\cite{Sch86,For96}) that include state-dependent correlations derived
from bare nucleon-nucleon interactions.  Experiments on complex
(A$>$4) nuclei have been performed \cite{Lap93} abundantly, but
theoretical calculations (Green's function \cite{Geu96} and cluster
VMC \cite{VNeck98}) are more difficult, and have been limited to a few
closed-shell nuclei; as discussed below, these usually overpredict the
normalization of the cross sections.  Typically, these data
(consisting mainly of knockout of valence protons) are analyzed by
comparing to calculations based on mean-field theory (MFT) that do not
include (short-range and other) correlations, and by identifying the
required renormalization as the spectroscopic factor.  The resulting
factors \cite{Lap93,Pan97} (about 60-65\% for nuclei
ranging from A=6 to 209), are usually interpreted as evidence for the
presence of important correlations.

A quenching of this kind was predicted for infinite nuclear matter
\cite{BenF89,RamD91} but the extension of this result to finite nuclei
is not straightforward due to the coupling to surface vibrations that
affects the strength for transitions near the Fermi edge.  For the
nucleus $^{16}$O the effect of both short and long-range correlations
was calculated with a Green's function method \cite{Geu96} resulting
in a reduction of the strength to 0.76 of the MFT value.  However, the
inclusion of center-of-mass effects will probably increase this value
to $\sim$ 0.81 \cite{VNeck98}, which is still considerably larger than
the observed \cite{Leu94} $1p$ strength ($\sim$0.6) at small
excitation energies.
It may be that $^{16}$O is an exceptionally difficult case;
more success has been gained with the larger nuclei $^{48}$Ca and
$^{90}$Zr \cite{Rij96}, although these calculations must use a G-matrix
representation of the NN force, a step away from the bare interaction.
In the present case, the VMC method uses the bare interactions
to produce rather sophisticated six- and seven-body wave functions,
including strong state-dependent correlations, which show the clustering
expected in light p-shell nuclei.

The theoretical description of the reaction ($e,e'p$) has been
given in detail elsewhere \cite{DieF75}. In Plane-Wave Impulse
Approximation (PWIA), the expression for the cross section reads:

\begin{eqnarray}
{d\sigma \over dE_{e'}d\Omega_{e'}dT_{p}d\Omega_{p}} =
K\sigma_{ep}S(E_{m},{\bf p}_{m}),
\label{eq:sigma}
\end{eqnarray}
where the spectral function $S(E_{m},{\bf p}_{m})$ denotes the
probability to find a proton with separation energy and momentum
$(E_{m},{\bf p}_{m})$ in the nucleus.  The quantity $K\sigma_{ep}$ is
the product of a phase space factor and the elementary off-shell
electron-proton cross section \cite{Fore83} that describes the
coupling of the virtual photon to the proton.  In this paper we
consider the transitions to the $0^{+}$ ground state ($E_{m}$=9.97
MeV) and $2^{+}$ first excited state ($E_{m}$=11.77 MeV) in $^{6}$He.
Hence in Eq.  (1) $E_{m}$ has two discrete values and we can obtain
the momentum distribution $\rho({\bf p}_{m})$ for each transition by
integrating $S(E_{m},{\bf p}_{m})$ over the appropriate peak in
$E_{m}$.  In PWIA the momentum distributions are related to the
overlap wave function $\langle^{6}$He$\mid~a({\bf
r})\mid^{7}$Li$\rangle$ via

\begin{eqnarray}
\rho({\bf p}_{m}) &=& \mid \int e^{i {{\bf p}_{m}\cdot {\bf r}}}
\langle {\rm ^{6}He}\mid a({\bf r}) \mid{\rm ^{7}Li}\rangle {\rm d}{\bf r}
\mid^{2}.
\label{eq:rho}
\end{eqnarray}
For the overlap wave functions we take either the MFT or the VMC
results, which are discussed below.  In order to account for Coulomb
distortion of the electron wave functions and for final-state
interaction (FSI) between the outgoing proton and the residual
$^{6}$He nucleus we use a CDWIA procedure \cite{GiuP88}.  Here the FSI
are treated via an optical-model potential, the parameters of which
were taken from a 100 MeV proton scattering experiment on $^{6}$Li
\cite{LiM68}.  For the extrapolation of these parameters to 90 MeV
protons and $^{6}$He we used Schwandt's \cite{Sch82} global
parameterization for a large number of nuclei and energies.  The
uncertainty due the treatment of the FSI was estimated by also
extrapolating the optical-model parameters from proton scattering data
at lower \cite{Bra72} and higher \cite{Hen81,Moa80} energies.  This
yielded model uncertainties on the spectroscopic factors of 6\%
and on the deduced rms radii of the overlap wave functions of 2\%.

\begin{figure}
\vspace{0.1 cm}
\centerline{\epsfysize=7.5cm \epsfbox{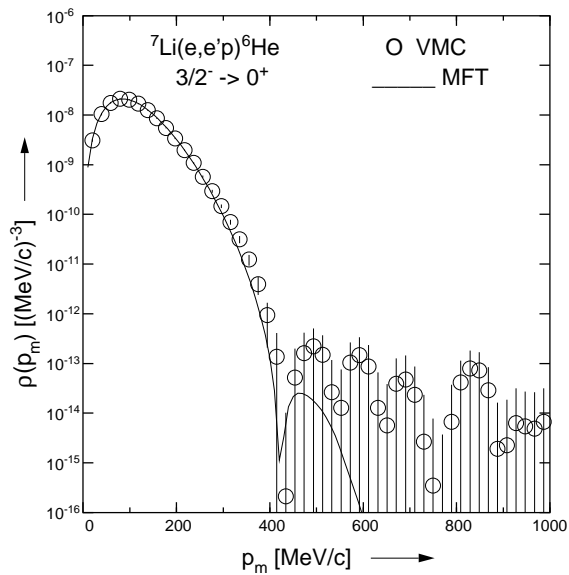} }
\vspace{0.4 cm} \caption[ ]{ Momentum distributions in PWIA for the
ground-state overlap wave function
$\langle^{6}$He$\mid~a({\bf r})\mid^{7}$Li$\rangle$ as calculated in
MFT (solid curve) and in VMC
(circles).  The error bars on the VMC data denote the uncertainty due
to Monte Carlo sampling.  The normalization of the MFT wave function
has been chosen identical to that of the VMC one.}
\label{gsr.gs}
\end{figure}

In mean-field theory the overlap wave function
$\langle^{6}$He$\mid~a({\bf r})\mid^{7}$Li$\rangle$ reduces to a
single-particle wave
function ${\phi_{\alpha}({\bf r}})$ since it is the product of two
Slater determinants. It is taken as the solution of the
Schr\"{o}dinger equation with a Woods-Saxon potential that reproduces
the appropriate binding energy. The radius of the potential is chosen
such that the calculated momentum distribution fits the experimental
data. Based on the VMC calculations, which predict dominant
$1p_{3/2}$ and $1p_{1/2}$ amplitudes for the two transitions,
respectively, we choose a MFT $1p_{3/2}$ wave function for the
$3/2^{-} \rightarrow 0^{+}$ ground state transition and a $1p_{1/2}$
MFT wave function for the $3/2^{-} \rightarrow 2^{+}$ transition to
the first excited state.

Quantum Monte Carlo calculations of the ground and low-lying excited
states for six- and seven-body nuclei have been made \cite{Pud97}
using a realistic Hamiltonian containing the Argonne $v_{18}$
two-nucleon \cite{Wir95} and Urbana IX three-nucleon \cite{Pud95}
potentials (AV18/UIX model). These calculations start with trial functions,
$\Psi_V(J^{\pi},T)$, constructed from products of two- and three-body
correlation operators acting on a fully antisymmetrized set of
one-body p-shell basis states that are $LS$ coupled to the specified
quantum numbers. Metropolis Monte Carlo integration \cite{Wir91} is used to
evaluate $\langle\Psi_V\mid H\mid\Psi_V\rangle$ and diagonalize in
the one-body p-shell basis, giving upper bounds to the energies of
these states. The trial functions are then used as input to the
Green's function Monte Carlo (GFMC) algorithm, which projects out
excited state contamination in the trial function by means of
the Euclidean propagation $\Psi(\tau) = \exp [ - ( H - E_0) \tau ]
\Psi_V$. The GFMC results are believed to be within $\sim$ 1\% of the
exact binding energy for the given Hamiltonian.

For the AV18/UIX model, the GFMC energy for the $^7$Li ground state
is -37.4(3) MeV, where the number in parentheses is the statistical
error due to the Monte Carlo energy evaluation.
The $^6$He(0$^+$) ground state is at -27.6(1) MeV and the $^6$He(2$^+$)
excited state is at -25.8(2) MeV.
While these {\it ab initio} energies are about 5\% above experiment (which
we attribute to inadequacies of the AV18/UIX model) the relative
excitations of 9.8(3) and 11.6(3) MeV are fairly close to experiment.
The VMC energies are not as good, but the one- and two-body VMC and
GFMC density distributions are very similar, giving us some
confidence in using the $\Psi_V$ to study reactions.
Recent calculations using the equivalent $\Psi_V$ for $^6$Li gave a very
satisfactory description of both elastic and inelastic electron scattering
form factors \cite{Wir98b}. For the present work, the
$\langle^{6}$He$\mid~a({\bf r})\mid^{7}$Li$\rangle$ overlaps
were calculated using the techniques of Refs.\cite{Sch86,For96}.

\begin{figure}
\vspace{0.1 cm}
\centerline{\epsfysize=8.0cm \epsfbox{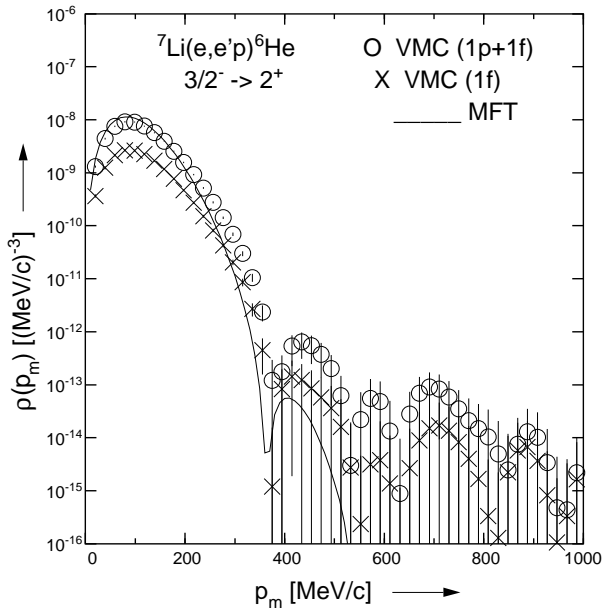} }
\vspace{0.4 cm} \caption[ ]{Same as Fig. \ref{gsr.gs} but for the
transition to the first excited state. The MFT wave function is pure
$1p_{1/2}$, the total VMC wave function (circles) contains $1p$ and
$1f$ (crosses) components.}
\label{esr.gs}
\end{figure}

In Figures \ref{gsr.gs} and \ref{esr.gs} we compare the plane wave
MFT and VMC wave functions in momentum space out to large momenta. In
order to facilitate the comparison we have scaled the MFT overlap
wave functions such that their normalization is identical to that of
the VMC wave functions. For the ground state transition (see Fig.
\ref{gsr.gs}) we observe that the MFT and VMC wave function are
practically identical up to $p_{m}$ = 400 MeV/c, whereas above this
momentum the VMC wave function is appreciably larger due to the
inclusion of short-range and tensor correlations, which are absent in MFT.
The transition to the first excited state contains both $1p$ and $1f$
components, as shown in Fig. \ref{esr.gs}. Here the deviation between
the VMC and MFT wave function already starts at 250 MeV/c because in
MFT the wave function is purely $1p_{1/2}$, whereas the VMC overlap
contains four components ($1p_{1/2}$, $1p_{3/2}$, $1f_{5/2}$,
$1f_{7/2}$). In addition to the effect of correlations
these extra components cause an appreciable enhancement of the VMC
wave function at high momentum relative to the MFT wave function.

The experiment was performed with the 1\% duty factor electron beam
from the NIKHEF medium-energy accelerator (MEA) and the
high-resolution two-spectrometer setup in the EMIN end station
\cite{Vri84}.  The data were taken concurrently with those for the
reaction $^{32}$S($e,e'p$) \cite{Wes92,Wes97} for which purpose a
self-supporting disc of Li$_2$S was used as a target (thickness
roughly 25 mg/cm$^2$).  The target could withstand maximum average
currents of 6 $\mu$A when rotated continuously.  The target thickness
was monitored via frequent measurements of elastic scattering.  The
measurements were carried out in parallel kinematics for an outgoing
proton energy of 90 MeV. As a result we needed two incident energies
(329.7 and 454.7 MeV) to cover the missing momentum range of -70 to
260 MeV/c.  Since the beam was tuned in dispersion matching mode
\cite{LapW80} we could achieve an $E_{m}$ resolution of 180 keV
(FWHM), sufficient to separate the discrete transitions from the two
reactions.

\begin{figure}
\vspace{0.1 cm}
\centerline{\epsfysize=7.5cm \epsfbox{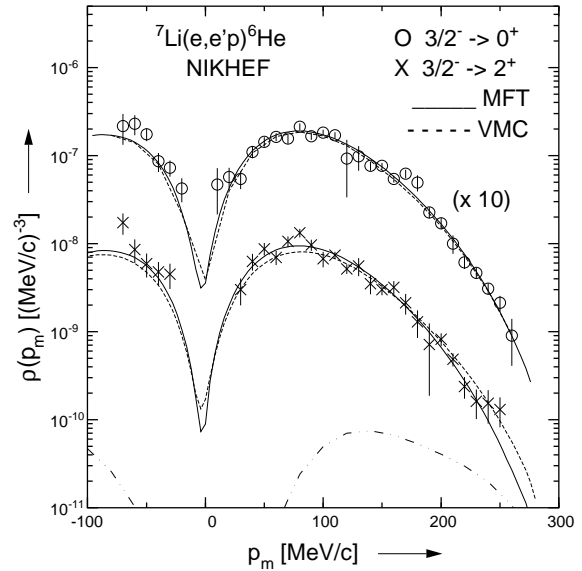} }
\vspace{0.4 cm} \caption[ ]{Experimental momentum distributions
for the transitions to the ground state (circles) and first
excited state (crosses) in the reaction
$^{7}$Li($e,e'p$)$^{6}$He, compared to CDWIA calculations with MFT
(solid) and VMC (dashed) wave functions. The
dash-dot-dot curve represents the $1f$ contribution to the full
VMC curve for the transition to the 2$^{+}$ state. The
error bars on the data are statistical only. For clarity data and
curves for the ground state transition have been scaled by 10.}
\label{mom.ps}
\end{figure}

The data analysis was performed in a standard way described in detail
elsewhere \cite{HerB88}. From the measured cross sections we
determined momentum distributions by integrating over the appropriate
missing-energy peak and by dividing out $K\sigma_{ep}$, for which we
used the current-conserving expression $\sigma_{ep}^{cc1}$ of De
Forest \cite{Fore83}. The resulting experimental momentum
distributions are shown in Fig. \ref{mom.ps},
where only the statistical errors are shown. The experimental
systematic uncertainty on these data is 5\%.

In the only one earlier reported \cite{NakH78} study of the reaction
$^{7}$Li($e,e'p$)$^{6}$He the missing-energy resolution of 7 MeV was
insufficient to separate the two transitions presented in this Letter.
However, when corrected for the
\begin{table}

\caption{Spectroscopic factors ($S$) and rms-radii deduced in the
present experiment for the transitions to the 0$^+$ and
2$^+$ state in $^{6}$He (first row).  The listed
errors include statistical, systematic and model uncertainties.  The
second (third) row presents the corresponding values for the VMC
calculation with $1p$ (1p+1f) wave function components.}
\begin{tabular}{lllllll}
Model & $S$ & $S$ & $S$ & rms[fm] & rms[fm]\\
      & 0$^+$ & 2$^+$ & 0$^+$+2$^+$ & 0$^+$ &2$^+$\\ \hline
Exp. ($1p$)&0.42(4)&0.16(2)&0.58(5)&3.17(6)&3.47(9)\\
VMC ($1p$)&0.41 &0.18 &0.59 &3.16 &3.14\\
VMC ($1p+1f$)&0.41 &0.19 &0.60 &3.16 &3.16\\
\end{tabular}
\label{table1}
\end{table}
\noindent presence of some unresolved $1s$
knockout strength and the difference in ejected proton energy, their
momentum distribution, integrated over the region $E_{m}$=6-15 MeV,
agrees within error bars with that for the sum of the two transitions
studied here.

In order to compare the theoretical calculations with the data we
carried out CDWIA calculations with the MFT and VMC wave functions as
input.  For the mean-field calculations we treated the normalization,
i.e., the spectroscopic factor $S$, and the radius of the WS potential
(that fixes the rms radius $<r^{2}>^{1/2}$ of the wave function) as
free parameters to be determined from a least squares fit to the data.
The resulting values are listed in table 1.  The summed spectroscopic
strength for $1p$ knockout is 0.58$\pm$0.05, where we have included
the experimental systematic uncertainty and the uncertainty due to
the choice of the optical potential.
The observed reduction of the single particle strength to 58\% of the
MFT value (which is unity for a single proton in the $1p$ shell) is in
good agreement with the reduction found for a large number of other
complex nuclei \cite{Lap93}.

Figure \ref{mom.ps} also shows the calculated momentum distributions
with the VMC wave functions, which are essentially parameter free.
The agreement with the data is very good as shown in table 1 where the
calculated spectroscopic factors with these wave functions are given.
The summed strength (0.60) for both transitions agrees within error
bars perfectly with the value 0.58$\pm$0.05 deduced from the MFT
analysis.

The VMC rms radius for the ground state transition agrees with the
value deduced from the MFT analysis, showing that the calculated
VMC ground state wave functions for $^{6}$He and
$^{7}$Li have the correct shape. For the transition to the first
excited state the rms radius of the VMC wave function is smaller than
that found in the MFT analysis. This is caused by the different
structure for both overlaps: the MFT wave function was assumed to be
pure $1p_{1/2}$, whereas the VMC wave function contains $1p_{3/2}$,
$1f_{5/2}$ and $1f_{7/2}$ components in addition. The contributions
of the $1f$ components, which depend sensitively on the details of
the nucleon-nucleon interaction employed, would show up in
measurements at higher $p_{m}$ and could thereby serve as a further
accurate test of the VMC wave functions.

In summary we conclude that
for the first time structure calculations for a complex nucleus, based
on a realistic nucleon-nucleon force, have been performed and compared
to (new) experimental data for the reaction $^{7}$Li($e,e'p$).  The
calculated spectroscopic strength (0.60) explains the reduction of the
strength to 0.58$\pm$0.05 found in a MFT analysis of the data, while
the calculated shape of the momentum distributions for $1p$
transitions nearly coincides with the experimental data.
Thus we have confirmed the necessity of including full correlations in
the nuclear wave functions.

We thank Dr. G.~van der Steenhoven for a helpful discussion.
This work is
part of the research program of the Foundation for Fundamental
Research of Matter (FOM), which is financially supported by the
Netherlands' Organisation for Advancement of Pure Research (NWO). The
work of R.B.W.\ is supported by the U.S.\ Department of Energy,
Nuclear Physics Division, under Contract No. W-31-109-ENG-38.

\end{document}